\documentclass[seceq]{ptptex}
\usepackage{type1cm, 
hyperref}

\usepackage{graphicx}

\newcommand{\la}{\lambda}

\DeclareMathOperator*{\Tr}{{\rm Tr}}

\newcommand{\al}{\alpha'}

\newcommand{\Fcal}{{\cal F}}

\newcommand{\thet}{\theta_k}

\title{
Two Polyakov Loop Correlators\\
\vskip 0.1 truecm
from D5-branes at Finite Temperature}

\author{
Ta-Sheng {\sc Tai}\footnote{E-mail: tasheng@hep-th.phys.s.u-tokyo.ac.jp}}

\inst{
Department of 
Physics, University of Tokyo, Tokyo 113-0033, Japan}

\abst{
We study two Polyakov loop correlators in large $N$ limit of ${\cal{N}}=4$ super 
Yang-Mills theory at finite temperature using 
the AdS-Schwarzschild black hole. In the case that one of the two loops is of 
the anti-symmetric 
representation, we use D5-branes to evaluate them. 
The phase structure of these correlators is also examined. 
A previous result, derived in hep-th/9803135 and hep-th/9803137, is realized 
as a limiting case.}

\newpage

\begin{document}

\maketitle

\section{Introduction and summary}
According to the pioneering works in Refs. 1) and 2), in the 
large 't Hooft coupling regime, 
two Wilson loop correlators or two Polyakov loop correlators 
have been 
computed with the Nambu-Goto action of a classical open string 
in the dual geometry.$^{3)-5),~18)}$ 
Gross and Ooguri pointed out in Ref. 5) that there exists a 
phase transition in this kind of correlator, which 
is best understood within the classical 
supergravity picture. This is because the topology of the open 
string worldsheet is an annulus, and 
the supergravity approximation breaks down when the circle of the 
annulus is of stringy size, $\simeq l_{s}$. 
Beyond that point, the supergraviton exchange between two loops becomes 
a more suitable description in the bulk. 
In other words, this change of topology gives rise to a phase transition.

Recently, due to the great progress made in 
our understanding of the holographical correspondence 
between 
Wilson loops of 
higher representations and D-branes,$^{7)-15)}$ 
to realize the correlator using a D-brane/F-string system is thus natural. 
For example, at zero temperature, by making use of D-branes, 
the anti-symmetric-anti-symmetric 
and fundamental-anti-symmetric 
Wilson loop correlators 
are studied in Refs. 16) and 17), respectively, following 
the early works.$^{18)-22)}$

In this paper, we study the finite temperature fundamental-anti-symmetric Polyakov 
loop correlators, 
where the loops wind around the temporal circle. 
By constructing the D5-brane solution in the dual AdS-Schwarzschild black hole, 
we are able to compute the large $N$ limit of the correlator, realized as 
the NG action of a macroscopic string stretching between 
the very D5-brane and the AdS boundary. 
Since this D5-brane wraps an $S^4$ of the internal $S^5$, 
we find that the angular dependence in $S^5$ plays a role in the 
Coulomb-like 
behavior of the correlator, for small separation between two loops.

We also find that the string worldsheet remains in the annulus phase 
even if a maximal separation is reached; 
that is, the Gross-Ooguri transition does not take place. 
In addition, 
the known result presented in Ref. 3) and 4) is obtained by examining the 
limiting behavior of our solution.

The outline of this paper is as follows. 
In $\S 2$, the AdS-Schwarzschild black hole is reviewed. 
In $\S 3$, we construct the D5-brane solution embedded in the 
dual black hole geometry. In $\S 4$, we evaluate two 
Polyakov loop correlators using 
D5-brane solutions. In $\S 5$, comments on 
the Gross-Ooguri transition and limiting cases are given. 
\section{AdS-Schwarzschild black hole}

We first describe the geometry of the five-dimensional 
AdS-Schwarzschild black hole, 
which is used to 
study the large $N$ boundary field theory, i.e. ${\cal{N}}=0$ four-dimensional 
SU($N$) Yang-Mills on $S^1 \times R^3$. At high temperature, 
it effectively becomes $\text{QCD}_3$, because of the thermal circle $S^1$
 (see Refs. 5) and 6)). 
In order to obtain the dual geometry, one can apply a 
Wick rotation to the Type IIB classical solution of $N$ 
coincident non-extremal D3-branes. 
What is of interest is its near horizon 
geometry, which is just the AdS-Schwarzschild black hole times an $S^5$. 
The metric is 
\begin{align}
\begin{aligned}
 \frac{ds^2}{\al}&=\frac{r^2}{R^2}(f(r)dt^2 + \sum^3_{i=1}dx_i^2)+R^2 f(r)^{-1}\frac{dr^2}{r^2}
 +R^2(d\theta^2 +\sin^2\theta d\Omega_4^2),\\
 f(r)&=1-\frac{r_0^4}{r^4}, ~~~~R^2=\sqrt{4\pi g_s N}=\sqrt{\lambda}, 
~~~~r_0=\pi\sqrt{4\pi g_s N T_H^2}.
\label{me1}
\end{aligned}
\end{align}
Despite the presence of $S^5$, the black hole has the topology of $D_2\times R^3$. 
To keep the near horizon manifold smooth, without any 
conical singularity, the Euclidean time direction is periodically 
identified as $t\sim t+\frac{1}{T}$, 
where $T_H=T/\pi$ is the Hawking temperature. 
In Refs. 3) and 4), by use of the dual AdS black hole, 
the quark-anti-quark potential $V_{q \bar q}$ in large $N$ limit is 
derived via 
the correlator of 
two temporal Wilson (Polyakov) loops, defined as 
\begin{align}
\begin{aligned}
\langle P(0) P(d)\rangle\simeq e^{-V_{q \bar q}(d,T)/T}=e^{-S_{min}},\\
P=\frac{1}{N}\Tr U,~~~~~~~U:=P~e^{i\int_0^{1/T} dtA_0},
\end{aligned}
\end{align}
where $d=|\triangle \vec{x}|$ is 
the separation between two loops in $R^3$. 
As found in Refs. 3) and 4), at low temperature (i.e. for $d\ll1/T$), there exists a minimal 
surface bounded by two Polyakov loops, 
which dominates the connected correlator. 
Contrastingly, at high temperature (i.e. for $d\gg1/T$), 
the inter-potential vanishes, due to the thermal screening.

In this paper, 
we consider another type of connected correlation function, 
\begin{align}
F(\thet,d,\Theta)=\frac{\langle P_k(0,\thet){P_{\square}}(d,\Theta) \rangle
-\langle P_k(0,\thet)\rangle
\langle {P_{\square}}(d,\Theta) \rangle}{\langle P_k(0,\thet)\rangle},
\label{cor}
\end{align}
where the subscript $k$ $({\square})$ denotes the $k$-th anti-symmetric 
(fundamental) 
representation. The quantity 
\eqref{cor} 
is realized in the dual geometry as a string minimal surface stretching between 
the AdS boundary and a D5-brane. This is because 
the vev $\langle P_{k}(0,\thet)\rangle$ is now 
related to a D5-brane carrying $k$ F-string charge, 
which probes the black hole geometry \eqref{me1} and is wrapped on an $S^4$ at 
$\theta=\thet$. 
In addition, the angular dependence $\Theta$ in \eqref{cor} arises from the fact that 
generally two loops are separated in 
$S^5$, e.g. 
$P(d,\Theta)$ is located at a point $n^I=(\cos\Theta, \sin\Theta,0,0,0,0)$ in $S^5$, 
where $n^I$ denotes the six-dimensional unit vector.

\begin{figure}
\centerline{\scalebox{0.4}{\includegraphics{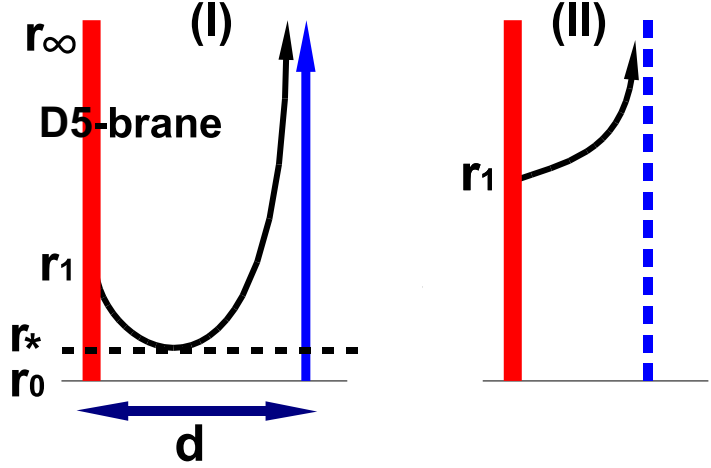}}}
\caption{(I) A schematic picture of the annulus phase (the J-shaped string) 
with its tip at $r_\ast$ in the case $0\le\thet\le\frac{\pi}{2}$. 
Because the D5-brane is much 
heavier than the fundamental string, it is not pulled. 
$d$ denotes the separation between 
two Polyakov loops. The disk phase (blue arrow) is also depicted on the RHS. 
(II) The annulus phase depicted in the case 
$\frac{\pi}{2}<\thet\le{\pi}$. Note that the slope at $r_1$ is determined by 
$\cos\thet$ as in \eqref{bo1}.}
\label{1}
\end{figure}

In large $N$ limit, $\langle P_{\square}(d,\Theta) \rangle=e^{-S_{min}}$ has been 
derived 
by using a fundamental string, which 
extends over $(t,r)$ with constant 
$(x_i,\Omega_5)$ coordinates. 
From the Nambu-Goto action (in the static gauge $\tau=t$), 
\begin{align}
\begin{aligned}
S_{min}=\frac{1}{2\pi\al}\int d\tau d\sigma\sqrt{det~G},
\label{NG}
\end{aligned}
\end{align}
where $G$ is the induced metric, it is found that 
\begin{align}
S_{min}=\frac{1}{{2\pi}T}(r_{\infty}-r_0). 
\label{disk}
\end{align}
The above string configuration is referred to as the $disk$ phase, which is 
compared with the $annulus$ phase below (see Fig. \ref{1}).

\section{D5-brane embedding}
We now construct the D5-brane solution which is expected to be dual to 
the Polyakov loop of the $k$-th anti-symmetric 
representation on the boundary.

By rescaling as $(t,x_i)\rightarrow {R^2}(t,x_i)$, 
the metric \eqref{me1} can be rewritten as 
\begin{align}
{ds^2}=L^2\big(r^2 f(r)dt^2 + \frac{dr^2}{r^2 f(r)} + r^2 dx_i^2 + d\theta^2 + 
\sin^2\theta d\Omega_4^2 \big), ~~~L^2=\al\sqrt{\lambda}.
 \label{me2}
\end{align}
Then, using the Euclidean DBI action and the RR four-form 
\begin{align}
\begin{aligned}
S_{bulk}&=T_5 
\int d^6\xi ~e^{-\phi}\sqrt{det(G+{\cal F})}-iT_5\int {\cal F}\wedge C_4, \\
C_4&=L^4 (\frac{3}{2}\thet-\sin 2\thet+\frac{1}{8}\sin4\thet)~\text{vol} S^4,
\end{aligned}
\end{align}
as well as the ansatz (in the static gauge) for the D5-brane 
\begin{align}
(t,r,x_i,\theta)=\big(\xi^0,r(\xi^1),0,\theta_k\big), ~~~~
\label{ansatz}
\end{align}
we arrive at 
\begin{align}
\begin{aligned}
S_{bulk}&=\int d\xi^0 d\xi^1 {\cal L},\\
{\cal L}&=T_5\frac{8\pi^2}{3}L^4 \Big(\sin^4 \thet\sqrt{L^4 r'^2 +{\cal F}^2}
-i{\cal F}
(\frac{3}{2}\thet-\sin 2\thet+\frac{1}{8}\sin4\thet)\Big),
\end{aligned}
\end{align}
where the prime denotes differentiation w.r.t. $\xi^1$, and 
we have integrated out the unit $S^4$ volume $(=\frac{8\pi^2}{3})$. 
The e.o.m. of $\Fcal=\Fcal_{01}$ is 
\begin{align}
\frac{\delta {\cal L}}{\delta \Fcal}=-\frac{i}{2\pi \al}k
=T_5\frac{8\pi^2}{3}L^4\Big(\sin^4 \theta_k ~\Fcal 
(L^4 r'^2 +{\cal F}^2)^{{-1}/{2}}
-i(\frac{3}{2}\thet-\sin 2\thet+\frac{1}{8}\sin4\thet)\Big).
\end{align}
Choosing the gauge in which $r=\xi^1$, 
we note that $\Fcal=2\pi\al dA=-iL^2\cos\thet$. The F-string charge $k$ 
satisfies 
\begin{align}
k=\frac{2N}{\pi}(\frac{1}{2}\thet-\frac{1}{4}\sin 2\thet),
\label{k}
\end{align}
where, for small $\theta_k$, we have 
\begin{align}
k\simeq\frac{2N}{3\pi}\thet^3.
\label{small}
\end{align}
In summary, this D5-brane carries $k$ F-string charge and wraps spatially 
an $S^4$ as well as the $r$ direction from $r_0$ to $r_{\infty}$. 

\section{Annulus phase}
We now evaluate the annulus contribution. 
It is presented as a minimal 
surface bounded by the D5-brane constructed above and 
a fundamental Polyakov loop. Fig. \ref{1} presents a 
schematic picture of this configuration. 
We use techniques similar to those utilized in Refs. 3), 4) and 17) to 
obtain the solution and its corresponding on-shell action.

\subsection{Ansatz and equations of motion}

Inserting the 
ansatz for the string worldsheet 
\begin{align}
\begin{aligned}
t=\tau,\qquad r=r(\sigma),\qquad x_1\equiv x=x(\sigma),\qquad \theta=\theta(\sigma),
\label{ansatz2}
\end{aligned}
\end{align}we obtain 
the bulk Nambu-Goto action as 
\begin{align}
&S_{bulk}=\frac{\sqrt{\la}}{2\pi T}\int_{\sigma_1}^{\sigma_2} d\sigma 
\sqrt{r'^2 + r^4 f(r) x'^2+ r^2 f(r)\theta'^2},
\label{bulk-action}
\end{align}
where the prime denotes differentiation w.r.t. $\sigma$, and 
we have integrated out $\tau$.

At $\sigma=\sigma_1$, the worldsheet boundary is constrained to 
the rigid D5-brane and satisfies at least the following conditions: 
\begin{align}
x=0,\qquad \theta=\theta_k.
\label{bc1-1}
\end{align}
These are not all the conditions, however. 
Due to the D5-brane worldvolume gauge field excitation, 
we need to include another boundary term, 
\begin{align}
\begin{aligned}
 S_{bdy,1}=i\oint_{\sigma=\sigma_1} 
  dt ~A_{t} ~ +(\text{constant})=
 -i\frac{r(\sigma_1)-r_0}{2\pi\al T} \Fcal,~
    &&A_{t}=-\frac{r}{2\pi \alpha'}\Fcal=\frac{irL^2 \cos\theta_k }{2\pi \alpha'}.
 \label{gaugevol}
 \end{aligned}
 \end{align}
The term ``(constant)'' ensures that $S_{bdy,1}=0$ 
when $r_1=r(\sigma_1)$ reaches the horizon $r=r_0$, where 
the temporal circle of $t$ shrinks to a point, due to 
$g_{tt}(r_0)=0$.

Now we derive boundary conditions obeyed by the worldsheet using a variational principle. 
The variation of the bulk action at $\sigma_1$ gives boundary terms as 
\begin{align}
 \delta S_{bulk}|_{bdy,1}
=&-p_r \delta r-p_x \delta x-p_{\theta}\delta \theta,
\label{bv}
\end{align}
where the momenta are 
\begin{align}
\begin{aligned}
p_r=\frac{\sqrt{\la}}{2\pi T}\frac{r'}
{\sqrt{r'^2 + r^4 f(r) x'^2+ r^2 f(r)\theta'^2}},\\
p_x=\frac{\sqrt{\la}}{2\pi T}\frac{r^4 f(r)x'}
{\sqrt{r'^2 + r^4 f(r) x'^2+ r^2 f(r)\theta'^2}},\\
p_\theta=\frac{\sqrt{\la}}{2\pi T}\frac{r^2 f(r)\theta'}
{\sqrt{r'^2 + r^4 f(r) x'^2+ r^2 f(r)\theta'^2}}.
\label{p}
\end{aligned}
\end{align}
Further, the variation of $S_{bdy,1}$ is 
\begin{align}
\delta S_{bdy,1}=-\frac{L^2\cos\thet}{2\pi \al T }\delta r.
\end{align}Thus, 
we obtain totally 
\begin{align}
 \delta S_{bulk}|_{bdy,1}+\delta S_{bdy,1}
=(-\frac{\sqrt{\lambda}\cos\theta_k}{2\pi T}-p_r)
 \delta r- p_x \delta x- p_\theta \delta \theta=0. 
\end{align}
The second and third terms vanish because of 
\eqref{bc1-1}. The first term implies an 
additional boundary condition at $\sigma_1$, i.e. 
\begin{align}
-\cos\theta_k=\frac{r'}
{\sqrt{r'^2 + r^4 f(r) x'^2+ r^2 f(r)\theta'^2}}.
\label{bo1}
\end{align}

Now we solve the equations of motion subject to the above boundary
conditions. 
From \eqref{bulk-action}, the equations of motion for $x$ and $\theta$ read 
\begin{align}
\begin{aligned}
\frac{r^4 f(r)}{\sqrt{r'^2 + r^4 f(r) + r^2 f(r)\theta'^2}}=\alpha,\\
\frac{r^2 f(r)\theta'}{\sqrt{r'^2 + r^4 f(r) + r^2 f(r)\theta'^2}}=\beta,
\label{eom}
\end{aligned}
\end{align}
where we have fixed the reparameterization 
of $\sigma$ as $x=\sigma$. Here, 
$\alpha$ and $\beta$ are integration constants. We can choose $\alpha \ge0$
without loss of generality. Further, from \eqref{eom}, we have 
\begin{align}
\begin{aligned}
\theta'&=\gamma r^2, ~~~~~~~~~~\gamma=\frac{\beta}{\alpha},\\ 
r'&=\pm \sqrt{\frac{r^8 f(r)^2}{\alpha^2}-(1+\gamma^2)r^4 f(r)}.
\label{reom}
\end{aligned}
\end{align}
Therefore, the boundary condition \eqref{bo1} can 
be rewritten as 
\begin{align}
\frac{r_1'}{\sqrt{r_1^4 -r_0^4}}=\pm\sqrt{\frac{1+\gamma^2}{\tan\theta^2_k}}.
\label{bc111}
\end{align}
Note that the sign of $r'_1$ depends on $\cos\thet$ in accordance with 
\eqref{bo1}, and thus 
the $``+"$ branch in \eqref{bc111} corresponds to $\frac{\pi}{2}<
\thet\le\pi$, whereas the $``-"$ branch%
\footnote{We include $\thet=\frac{\pi}{2}$ in the $``-"$ branch.} corresponds to $0\le\thet\le\frac{\pi}{2}$. 
On the basis of \eqref{reom} and \eqref{bc111}, we can
explicitly express $r_1$ as
\begin{align}
r^4_1-r^4_0=(\alpha^2+\beta^2)\csc^2\thet.
\label{r1}
\end{align}
This implies that when $\thet\ll1$, $r_1$ is near $r_{\infty}$.

\subsection{$0\le\thet\le\frac{\pi}{2}$ case}

For $0\le\thet\le\frac{\pi}{2}$, \eqref{eom} suggests that 
\begin{align}
\begin{aligned}
\big[\frac{r'}{r^4 f(r)}\big]^2+\frac{1+\gamma^2}{r^4 f(r)}=
\frac{1+\gamma^2}{r_\ast^4 f(r_\ast)}=\text{constant}.
\label{eo}
\end{aligned}
\end{align}
In other words, $r_\ast$ is located where $\frac{dr}{d\sigma}=0$ 
(see Fig. \ref{1}) 
and is determined through \eqref{reom} 
as $r^4_\ast=r_0^4+(\alpha^2 +\beta^2)$. 
Then, from \eqref{eo}, we 
obtain
\begin{align}
\begin{aligned}
dx&=\pm\frac{\sqrt{b^4 -1} ~dy}{r_0\sqrt{(y^4-1)(y^4-a^4)}},\\
r_0\int^d_{0} dx&=\big(\int_a^{y_1} + 
\int_a^\infty\big) \frac{\sqrt{b^4 -1} ~dy}{\sqrt{(y^4-1)(y^4-a^4)}},
\label{e}
\end{aligned}
\end{align}
where we have chosen the proper branches in the second line 
and set 
\begin{align}
\begin{aligned}
y=\frac{r}{r_0}, ~~~~~~a=\frac{r_\ast}{r_0},~~~~~~\frac{b^4-1}{a^4-1}=\frac{1}{1+\gamma^2}.
\label{set}
\end{aligned}
\end{align}
As in Ref. 3), expressing \eqref{e} as an elliptical integral through 
a change of variable to
$2w(y)=(y^2/a + a/y^2)$, we arrive at 
\begin{align}
\begin{aligned}
d=\frac{1}{4r_0~\sqrt{2}}\sqrt{\frac{b^4-1}{a^3}}
\Big[(\int_{w(a)}^\infty + \int_{w(a)}^{w_1}) \frac{dw}{\sqrt{(w-1)(w^2-w(a)^2)}}\\
-(\int_{w(a)}^\infty + \int_{w(a)}^{w_1}) \frac{dw}{\sqrt{(w+1)(w^2-w(a)^2)}}
\Big],\\
\rightarrow~~~d=\frac{1}{4r_0~\sqrt{w(a)}}\sqrt{\frac{b^4-1}{a^3}}
\Big[K(\sqrt{\frac{w(a)+1}{2w(a)}})-K(\sqrt{\frac{w(a)-1}{2w(a)}})\\
+F(\chi_-,\sqrt{\frac{w(a)+1}{2w(a)}})-
F(\chi_+,\sqrt{\frac{w(a)-1}{2w(a)}})
\Big],
\label{Ell}
\end{aligned}
\end{align}
where 
\begin{align}
\chi_{\pm}=
\sin^{-1}\sqrt{\frac{w_1 -w(a)}{w_1 \pm1}}, ~~~~w_1=w(y_1), ~~~~
y_1=\frac{r_1}{r_0}.
\end{align}
Note that $F$ is the elliptical integral of the first kind, while 
$K$ is its complete 
counterpart. 
From \eqref{reom}, we can also integrate out the $\theta$ part, obtaining 
\begin{align}
d\theta=\pm\gamma r_0 \sqrt{b^4-1}\frac{y^2 dy}{\sqrt{(y^4-1)(y^4-a^4)}}.
\end{align}
Hence, we find 
\begin{align}
\begin{aligned}
\Theta-\thet=\gamma r_0 \sqrt{b^4-1}\frac{\sqrt{2a}}{8a}
\Big[(\int_{w(a)}^\infty + \int_{w(a)}^{w_1})\frac{dw}{\sqrt{(w-1)(w^2-w(a)^2)}}\\
+(\int_{w(a)}^\infty + \int_{w(a)}^{w_1})\frac{dw}{\sqrt{(w+1)(w^2-w(a)^2)}}
\Big],\\
=\gamma r_0 \sqrt{\frac{b^4 - 1}{8(a^2+1)}}
\Big[K(\sqrt{\frac{w(a)+1}{2w(a)}})+K(\sqrt{\frac{w(a)-1}{2w(a)}})\\
+F(\chi_-,\sqrt{\frac{w(a)+1}{2w(a)}})+
F(\chi_+,\sqrt{\frac{w(a)-1}{2w(a)}})
\Big],
\label{om}
\end{aligned}
\end{align}
where the branches are chosen as in the case of \eqref{Ell}. 
Note that \eqref{om} is invariant under 
$\Theta\to2\thet-\Theta$, $\beta\to-\beta$. In summary, 
as seen from \eqref{Ell} and \eqref{om}, $(d, \Theta)$ 
can be determined in terms of the three parameters $\thet,\alpha$ and $\beta$, 
if they exist.

Finally, 
the total energy $E$ 
is proportional to the on-shell action, 
$S_{tot}=(S_{bulk} +S_{bdy,1})$: 
\begin{align}
E=\frac{\sqrt{\lambda}}{2\pi}r_0
\Big[\big(\int_a^\Lambda+ \int_a^{y_1}\big) dy \sqrt{\frac{y^4 -1}{y^4 -a^4}}
- \cos\thet(y_1 -1)-(\Lambda -1)\Big],~~~~~\Lambda=\frac{r_\infty}{r_0}.
\label{ener}
\end{align}
Here, we have regularized $E$ by subtracting the term $(\Lambda -1)$. 
Note that $\frac{\sqrt{\lambda}}{2\pi}r_0(\Lambda -1)$ is precisely 
the disk phase energy 
mentioned in \eqref{disk}. 
The condition for the Gross-Ooguri transition to occur can be 
phrased as $E=0$, which means 
that the annulus phase has the same 
amount of energy as the disk phase.

\subsection{$\frac{\pi}{2}<\thet\le\pi$ case}
For $\frac{\pi}{2}<\thet\le\pi$, \eqref{r1} suggests that 
$r'$ in \eqref{reom} never becomes zero, unlike in the previous case, in which 
there exists a tip at $r_\ast$. In the present case, $r$ increases monotonically 
from $r_1$ to infinity [see (II) of Fig. \ref{1}]. 
According to the above computations, $d$, 
$\Theta$ and $E$ here can be expressed as follows: 
\begin{align}
\begin{aligned}
d=\frac{1}{4r_0~\sqrt{w(a)}}\sqrt{\frac{b^4-1}{a^3}}
\Big[K(\sqrt{\frac{w(a)+1}{2w(a)}})-K(\sqrt{\frac{w(a)-1}{2w(a)}})\\
-F(\chi_-,\sqrt{\frac{w(a)+1}{2w(a)}})+
F(\chi_+,\sqrt{\frac{w(a)-1}{2w(a)}})
\Big],\\
\Theta-\thet=\gamma r_0 \sqrt{\frac{b^4 - 1}{8(a^2+1)}}
\Big[K(\sqrt{\frac{w(a)+1}{2w(a)}})+K(\sqrt{\frac{w(a)-1}{2w(a)}})\\
-F(\chi_-,\sqrt{\frac{w(a)+1}{2w(a)}})-
F(\chi_+,\sqrt{\frac{w(a)-1}{2w(a)}})
\Big],\\
E=\frac{\sqrt{\lambda}}{2\pi}
r_0\Big[\int_{y_1}^\Lambda dy \sqrt{\frac{y^4 -1}{y^4 -y_1^4}}
- \cos\thet(y_1 -1)-(\Lambda -1)\Big].
\label{se}
\end{aligned}
\end{align}

\section{Comments}
\begin{figure}
\centerline{\scalebox{0.5}{\includegraphics{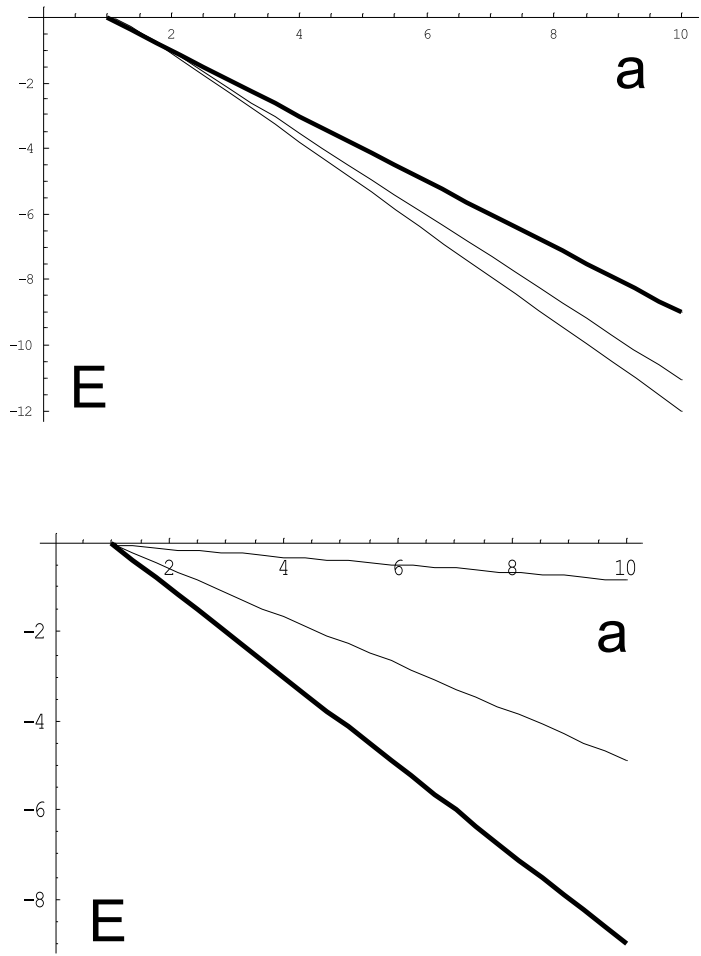}}}
\caption{(Upper panel) The $E$-$a$ relation in the case 
$0\le\thet\le\pi/2$ 
obtained from \eqref{ener}. Here, 
$\thet=\pi/2,~\pi/3$, and $\pi/4$ are plotted from top to bottom. 
(Lower panel) The case $\pi/2<\thet\le\pi$ obtained from \eqref{se}. Here, 
$\thet=9\pi/10,~2\pi/3$, and $\pi/2$ are plotted from top to bottom. 
Note that the profiles of $\thet=\pi/2$ in the two cases coincide.}
\label{2}
\end{figure}

We now attempt to obtain an intuitive understanding of 
the above solutions by studying their 
asymptotic behavior. 
In \eqref{Ell} and \eqref{se}, when $r_1,a\to \infty$ 
(i.e. $\alpha\to\infty$ with $\beta$ fixed), due to the relation 
$K(\sqrt{\frac{1}{2}})\simeq 2.086$ (finite) and the condition $\chi_{\pm}\to0$ 
(which allows us to ignore the elliptic integrals $F$), 
it is seen that $\Theta$ goes to $\thet$ for all given $\thet$, 
while the separation $d$ approaches zero in powers of $1/a$ as 
$\sim {1}/{a} +$ $\cal{O}$$({1}/{a^5})$.

In other words, at sufficiently low temperature (i.e. for $d\ll1/T$), the leading string energy 
$E(d)$ is proportional to $\cos^2(\frac{\thet}{2})\sqrt{\lambda}/d$ with 
$0\le\thet\le{\pi}$, which differs from the fundamental-fundamental case 
treated in Refs. 3) and 4) by the factor $\cos^2(\frac{\thet}{2})$.

In addition to the case $a=\infty$ taken care of above, 
$d$ also approaches zero asymptotically like%
\footnote{Here, the logarithmic divergence arises from the 
asymptotic expansion of $K(1+\epsilon)$.} 
$\lim_{\epsilon\to0}\big(\sqrt{\epsilon}\log\epsilon \big)\sim 0$ near the lower 
extreme, $a=1$. By noting that $d(a)$ is always positive for $1\le a\le\infty$, 
it is then straightforward to show that starting from zero separation 
(where $a=\infty$), we can reach an 
$a_M$ such that $d(a_M)=d_{max}$ is the maximum. 

Referring to Fig. \ref{2}, 
we see that for generic $\thet$, 
the phase transition occurs only at $a=1$, where 
$r_1=r_\ast=r_0$. This observation leads to the 
conclusion that by enlarging $d$ 
from zero, 
even when $d$ realizes its maximum at some $a=a_{M}>1$, 
the annulus phase still remains 
energetically favorable. This property differs from that in 
the fundamental-fundamental case, where the worldsheet splits at some $a_{M}\neq 1$.

It is also illuminating to check some limiting aspects of these 
solutions. We consider two situations: $\thet=0$ and $\frac{\pi}{2}$. 
As $k\rightarrow1,$%
\footnote{This serves as a 
non-trivial check in the sense that the D5-brane picture 
is valid only when $k$ is large, $\simeq{\cal{O}}(N)$.}
it is seen that 
$\lim_{k\rightarrow1}\theta_{k}\cong \left(\frac{3\pi}{2N}\right)^{1/3} 
\ll 1$ from \eqref{k}. 
Then, from the relation $\theta_k\ll 1$, we find that 
$r_1 \rightarrow \infty$ by 
virture of \eqref{r1}. 
Therefore, when $k=1$, we have
\begin{align}
E&=\frac{\sqrt{\lambda}}{2\pi T}
r_0\Big[2\int_a^\Lambda dy \sqrt{\frac{y^4 -1}{y^4 -a^4}}
-2(\Lambda -1)\Big],\label{ener2}\\
d&=\frac{1}{2r_0~\sqrt{w(a)}}\sqrt{\frac{b^4-1}{a^3}}
\Big[K(\sqrt{\frac{w(a)+1}{2w(a)}})-K(\sqrt{\frac{w(a)-1}{2w(a)}})\Big],\\
\Theta&=2\gamma r_0\sqrt{\frac{b^4-1}{8(a^2+1)}}
\Big[K(\sqrt{\frac{w(a)+1}{2w(a)}})+K(\sqrt{\frac{w(a)-1}{2w(a)}})\Big].
\end{align}
We thus see that the boundary term \eqref{gaugevol} 
provides exactly the subtraction term $(\Lambda -1)$, which 
can also be understood as a 
contribution from the denominator, $\lim_{k\to1}\langle P_k(0,\thet)\rangle$, 
of the correlator in \eqref{cor}. 
Note that \eqref{ener2} is simply the standard result 
derived in Refs. 3) and 4).

Next, in the case 
$\thet=\frac{\pi}{2}$, 
we see that $r_1=r_\ast$ from \eqref{r1}; i.e. 
the configuration is nothing but one-half of a full U-shaped string 
in the above $\thet=0$ case. In other words, 
its end on the D5-brane is now located at the tip where 
$\frac{dr}{d\sigma}=0$.

\subsection*{Acknowledgements}
We are grateful to Satoshi Yamaguchi for many valuable comments on this 
manuscript. 

\baselineskip=14pt

\end{document}